\documentclass[%
  twoside,
  preprint,
  amsmath,amssymb,
  aps,
  pra,
  nofootinbib,au
  showpacs,
  superscriptaddress,
  a4paper
]{revtex4-1}

\usepackage{graphicx}% 
\usepackage[usenames,dvipsnames]{xcolor}
\usepackage{siunitx}
\usepackage{subfigure}
\usepackage{enumitem}
\usepackage{subfigure}

\usepackage{tabularx}
\usepackage{booktabs}
\usepackage{multirow}

\usepackage{titlesec}

\usepackage{array}

\usepackage[utf8]{inputenc}
\usepackage[T1]{fontenc}

\usepackage{lipsum}

\newcommand{\BiCRCP}{BiCRCP~}

\graphicspath{{Figures/}{}}

\usepackage[
centering, includefoot,
text={7.1in,10.2in},
total={6.3in,8.75in},
top=0.8in, left=0.62in,
]{geometry}

\usepackage[
  bookmarks=false,
  colorlinks,
  linkcolor=blue,
  urlcolor=blue,
  citecolor=blue,
  plainpages=false,
  pdfpagelabels,
  final,
  breaklinks=true
]{hyperref}
\usepackage{natbib}
\makeatletter \def\NAT@def@citea{\def\@citea{\NAT@separator\,}} \makeatother

\newcommand{\orcid}[1]{%
  \href{%
    https://orcid.org/#1%
  }{%
   \,\protect\includegraphics[width=8pt]{ORCID-icon.png}%
  }%
}

\usepackage{physics}

\DeclareSIUnit{\au}{{a.u.}}

\newcommand{\nhphantom}[1]{\sbox0{#1}\hspace{-\the\wd0}}
\begin{document}

\title[Controlling polarization of attosecond pulses with plasmonic-enhanced bichromatic counter-rotating circularly polarised fields]
  {Controlling polarization of attosecond pulses with plasmonic-enhanced bichromatic counter-rotating circularly polarised fields}

\author{Irfana N. Ansari}
\affiliation{Department of Physics, Indian Institute of Technology Bombay,
            Powai, Mumbai 400076, India}

\author{Cornelia Hofmann}
\affiliation{Max Planck Institute for the Physics of Complex Systems,
N{\"o}thnitzer Stra{\ss}e 38, 01187 Dresden, Germany}

\author{Lukas Medi\v{s}auskas}
\affiliation{%
Max Planck Institute for the Physics of Complex Systems,
N{\"o}thnitzer Stra{\ss}e 38, 01187 Dresden, Germany}

\author{Maciej Lewenstein}
\affiliation{ICFO - Institut de Ciencies Fotoniques, The Barcelona Institute of Science and Technology, 
Av.  Carl Friedrich Gauss 3, 08860 Castelldefels (Barcelona), Spain}
\affiliation{ICREA, Pg. Llu\'is Companys 23, 08010 Barcelona, Spain}

\author{Marcelo F. Ciappina}
\email{marcelo.ciappina@gtiit.edu.cn}
\affiliation{ICFO - Institut de Ciencies Fotoniques, The Barcelona Institute of Science and Technology, 
Av.  Carl Friedrich Gauss 3, 08860 Castelldefels (Barcelona), Spain}
\affiliation{Institute of Physics of the ASCR, ELI Beamlines Project, Na Slovance 2, 18221 Prague, Czech Republic}
\affiliation{Physics Program, Guangdong Technion -- Israel Institute of Technology, Shantou, Guangdong 515063, China}
\affiliation{Technion -- Israel Institute of Technology, Haifa, 32000, Israel}

\author{Gopal Dixit}
\email{gdixit@phy.iitb.ac.in}
\affiliation{Department of Physics, Indian Institute of Technology Bombay,
            Powai, Mumbai 400076, India}

\date{\today}

\keywords{plasmonic fields, high-harmonic generation, bichromatic circularly polarised fields, attosecond pulses}

%%%%%%%%%%%%%%%%%%%%%%%%%%%%% Abstract: %%%%%%%%%%%%%%%%%%%%%%%%%%%%%%%%%%%%%%%%%%%%%%%
\begin{abstract}

The use of bichromatic counter-rotating laser field is known to generate high-order harmonics with non-zero ellipticity. 
By combining such  laser field  with a plasmonic-enhanced spatially inhomogeneous field, we propose a way to influence the sub-cycle dynamics of the high-harmonic generation process. Using the numerical solution of the time-dependent Schr{\"o}dinger equation combined with classical trajectory Monte Carlo simulations, we show that the change of the direction and the strength of the plasmonic field selectively enhances or suppresses certain recombining electron trajectories. This in turn modifies the ellipticity of the emitted attosecond pulses. 

\end{abstract}
%%%%%%%%%%%%%%%%%%%%%%%%%%%%%%%%%%%%%%%%%%%%%%%%%%%%%%%%%%%%%%%%%%%%%%%%%%%%%%%%%%%%%%%

\maketitle
%%%%%%%%%%%%%%%%%%%%%%%%%%%%%%%%%%%%%%%%%%%%%%%%%%%%%%%%%%%%%%%%%%%%%%%%%%%%%%%%%%%%%%%

\section{Introduction}

Control over the polarisation of attosecond pulses in 
extreme ultraviolet (XUV) and soft x-ray radiation is paramount 
to probing chiral-sensitive light-matter interactions such as 
x-ray magnetic circular dichroism~\cite{kfir2015generation, fan2015bright}, 
discrete molecular symmetries~\cite{reich2016illuminating, baykusheva2016bicircular, neufeld2019floquet},
magnetisation and spin dynamics~\cite{graves2013nanoscale, eisebitt2004lensless, radu2011transient, boeglin2010distinguishing}, and
recognising  chirality in molecules via photoelectron circular dichroism~\cite{cireasa2015probing, travnikova2010circularly, ferre2015table} at their intrinsic timescales.  
Following the proposal of Becker and co-workers~\cite{eichmann1995polarization, milovsevic2000attosecond, milovsevic2000generation}, a series of experiments have been
carried out to generate high harmonics with controlled polarisation~\cite{fan2015bright, kfir2015generation, chen2016tomographic, lambert2015towards, fleischer2014spin, ferre2015table, jimenez2017time, dorney2017helicity}. 
A combination of two counter-rotating circularly polarised 
laser pulses having fundamental ($\omega$) and its second harmonic ($2 \omega$) frequency is 
employed to generate circularly (or elliptically) polarised high-harmonics. 
The resultant electric field of $\omega$-$2 \omega$ combination exhibits 
trefoil symmetry and yields three radiation bursts per cycle of the fundamental field~\cite{eichmann1995polarization, milovsevic2000attosecond, milovsevic2000generation, medivsauskas2015generating, jimenez2017time}.  
The resulting high-harmonic spectrum contains doublets of circularly polarised harmonics with alternating helicity. The $3n+1$ and $3n+2$ harmonics follow the polarisation of
$\omega$- and $2\omega$-fields, respectively, whereas  $3n$ harmonics 
are parity forbidden~\cite{fleischer2014spin, alon1998selection, milovsevic2015high, neufeld2019floquet}. 

However, in High-Harmonic Generation (HHG) the 
control over the polarisation of the 
harmonics does not immediately 
entail the same control over the polarisation of underlying attosecond pulses.
Therefore, proposals have been put forward in recent years
to achieve the subcycle emission control by
either modifying the underlying medium, (e.g., by choosing a
suitable initial state of atoms~\cite{medivsauskas2015generating, milovsevic2015generation, 
ayuso2017attosecond, misha2018}, using different molecular systems~\cite{yuan2013single, 
mauger2014quantum}),  
or by modifying the laser driving schemes, (e.g., by
using non-collinear bichromatic counter-rotating 
circularly polarised laser pulses~\cite{hernandez2016schemes, hickstein2015non, huang2018polarization},
introduction of seed XUV pulse with suitable polarisation~\cite{dixit2018control}, 
tuning the ratio of the ellipticity and/or of the intensity of the bichromatic counter-rotating 
driving fields~\cite{jimenez2017time, dorney2017helicity, neufeld2018optical}), or by 
changing the time-delay between two driving 
pulses~\cite{jimenez2017time, frolov2018control}.
  
Since the experiment of Kim {\it et al.}~\cite{kim2008high}, a number of  
works have investigated the plasmonic-field assisted HHG 
using structured nano-objects~\cite{park2011plasmonic, pfullmann2014nano, han2016high, stebbings2011generation, sivis2013extreme, pfullmann2013bow, kruger2018attosecond, kruger2012attosecond, ciappina2017attosecond}.  
Moreover, different kinds of nanostructures such as 
metal nanotips~\cite{kruger2018attosecond, kruger2012attosecond}, 
metallic waveguides~\cite{park2011plasmonic}, nanoparticles~\cite{yang2013high} 
and plasmonic antennas~\cite{sivis2013extreme}, have been explored, while 
many more works tried
to explain HHG mediated by a plasmonic-enhanced field using numerical and theoretical 
methods~\cite{husakou2011polarization, ciappina2012high, ciappina2012enhancement, 
shaaran2012quantum,    yavuz2012generation, yavuz2013gas, shaaran2013quantum,  
shaaran2013high, perez2013beyond,  
fetic2013high, he2013wavelength, 
ebadi2014interferences, ciappina2014high,    luo2014efficient, chacon2015numerical,  feng2015attosecond,  ansari2018simultaneous}. 
Despite the abundance of research, most of the works on 
plasmonic-field assisted-HHG are investigating  
linearly polarised driving fields. 
In this regard, its not apparent  how the high-harmonic spectrum will 
behave when the linearly  polarised  pulse in the presence of plasmonic field  
is replaced by \BiCRCP fields. 
Will the pattern of circularly polarised doublets with alternating helicity be 
preserved throughout the spectrum? 
Will the increase in the energy cutoff be modified? 

\begin{figure*}
\includegraphics[width=.6\textwidth]{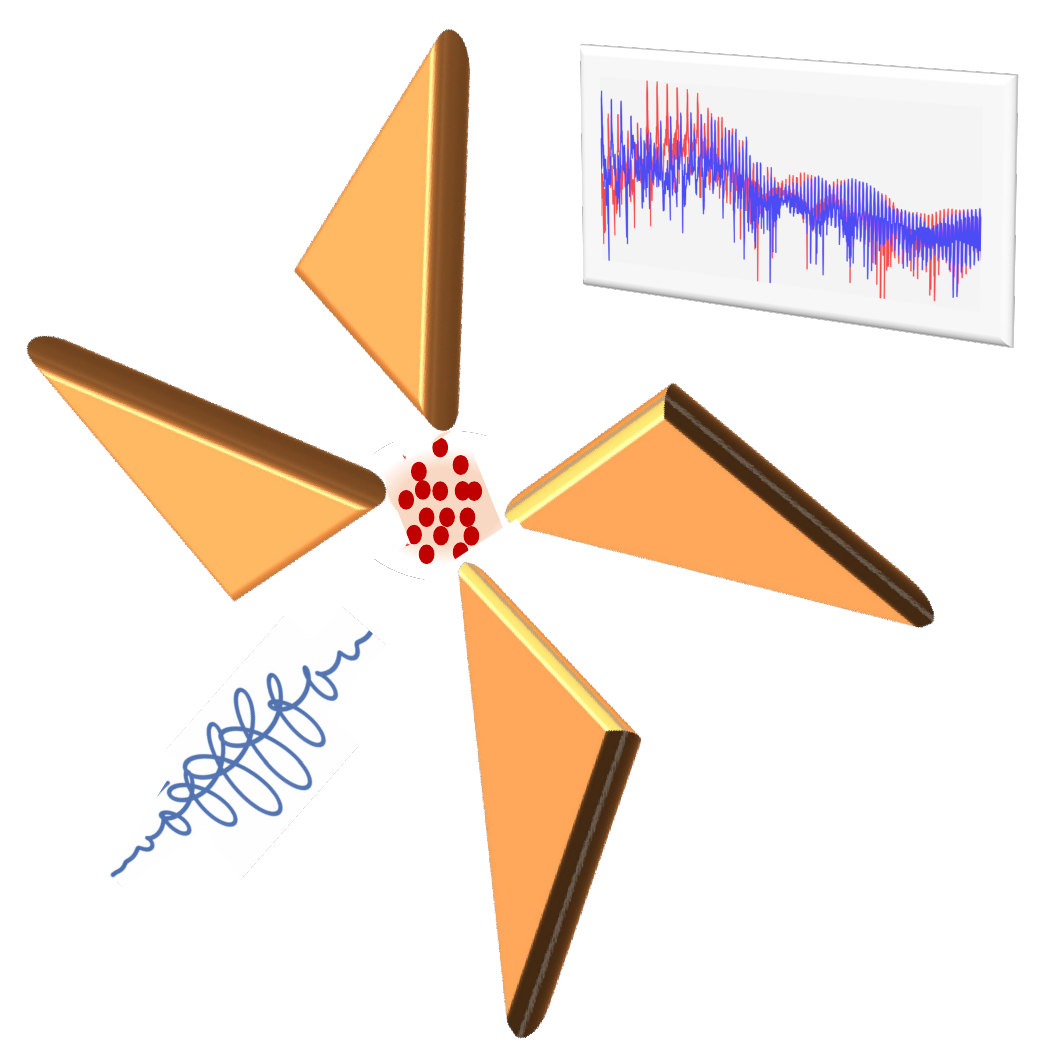}
\caption{Conceptual illustration of the HHG driven by plasmonic-enhanced bichromatic counter-rotating circularly polarised fields.
An incoming low intensity bicircular counter-rotating circularly polarised field (in blue), interacts with 
bowtie-shaped nano-objects and, 
as a result, a much higher intensity spatially nonhomogeneous laser field is generated, able to produce polarisation-resolved higher-order harmonics by the atoms located in the vicinity of the bow-tie elements.} \label{fig1}
\end{figure*}

In this work, we demonstrate, at the single atom level, the control over sub-cycle electron dynamics during HHG
by combining bichromatic counter-rotating circularly polarised driving fields (\BiCRCP) 
with plasmonic field enhancement in a 2D geometry (see Fig.~\ref{fig1}). 
The direction and strength of the plasmonic-enhanced field will provide 
the desired control over the sub-cycle electron dynamics, which 
in turn will allow us to exert influence over the attosecond pulses polarization. Atomic units are used throughout in the present work. 

\section{Theory}

In order to study HHG driven by plasmonic-enhanced \BiCRCP,
the time-dependent Schr{\"o}dinger equation (TDSE) is solved in two spatial dimensions (2D). In the present work we give a short summary and refer the reader, for more details, to Ref.~\cite{medivsauskas2015generating}. The 2D TDSE within length gauge can be written as 
\begin{equation}\label{eq01}
i \frac{\partial \Psi(\mathbf{r}, t) }{\partial t}  = [\hat{T} + V(\mathbf{r}) + \mathbf{r} \cdot  \mathbf{E}^{\mathrm{total}}(\mathbf{r} , t)  ] \Psi(\mathbf{r}, t)  
\end{equation}
where $\hat{T}$ is the kinetic energy operator in cartesian coordinates $(x,y)$ and  $\mathbf{E}^{\mathrm{total}}(\mathbf{r} , t) = \mathbf{E}^{\mathrm{hf}}(t)+\mathbf{E}^{\mathrm{pl}}(\mathbf{r}, t)$ is the total laser electric field, composed of the incoming laser electric spatially homogeneous field $\mathbf{E}^{\mathrm{hf}}(t)$ and the 
 plasmon-enhanced spatial nonhomogeneous part $\mathbf{E}^{\mathrm{pl}}(\mathbf{r}, t)$.  Equation (\ref{eq01}) is solved numerically for the soft-core potential given by
\begin{equation}\label{eq02}
V(\mathbf{r}) = -\frac{1+ 9 \, e^{-r^{2}}}{\sqrt{r^{2}+a}},
\end{equation}
where the parameter $a = 2.88172$
is used to describe a 2D model neon atom. The ionisation potential of the initial $p$-orbital then results to $I_p$ = 0.793 a.u. ($21.6$ eV)~\cite{barth2014numerical}. 
A Taylor-series propagator  with expansion up to eighth order is employed to time-propagate the 2D-TDSE on a cartesian grid~\cite{moler2003nineteen}. 
 We use a complex absorbing potential to avoid unphysical reflections from the boundary as
\begin{equation}
V_{\textrm{abs}}(\mathbf{r}) = \eta (\mathbf{r}-\mathbf{r}_{0})^{n},    
\end{equation}  
with $n = 3$, $\eta = 5 \times 10^{-4}$ and $\mathbf{r}_{0}$ is set to $\pm 70$ a.u. in both the $x$ and $y$ directions.
Finally, the temporal and spatial grid-step sizes are $dt$ = 0.005 a.u. and $dr$ = 0.2 a.u., respectively.

The incoming infra-red (IR) spatial homogeneous laser electric field has the following form:
\begin{equation}\label{eq03}
\mathbf{E}^{\mathrm{hf}}(t)  =   E_{0}^{i}~g(t)~[E_x(t)~\hat{x} + E_y(t)~\hat{y}],
\end{equation} 
where $E_{0}^{i}$  is the laser electric field strength and $g(t)$ is a trapezoidal envelope with five-cycle plateau and two-cycle rising and falling edges (in units of the fundamental laser frequency $\omega$). 
The incoming bichromatic driving  field consists  of the 
counter-rotating fundamental field (left-handed circularly polarised) and  the co-rotating 
second harmonic field (right-handed circularly polarised). The components 
$E_x(t) = \cos(wt) + \cos(2wt)$ and 
$E_y(t) = \sin(wt) - \sin(2wt)$ are the $x$- and $y$-components of the total driving IR field, respectively,
where $\omega = 0.05$ a.u., corresponding to a laser wavelength $\lambda\approx900$ nm.

At nanometric scale, the plasmon-enhanced electric field becomes space-dependent, which significantly alters the electron dynamics. 
The 2D plasmonic-enhanced spatial nonhomogeneous field is, then, described as 
\begin{equation}\label{eq04}
\mathbf{E}^{\mathrm{pl}}(x, y, t) = E_0~g(t)~\left[\beta_{x}~h(x)~E_{x}(t)~\hat{x} 
           + \beta_{y}~h(y)~E_{y}(t)~\hat{y}\right].
\end{equation}
In our study we set $E_0=0.05$ a.u., that corresponds to a laser intensity $I=E_0^2\approx 9 \times 10^{13}$ W/cm$^2$, large enough as to generate HHG out of a set of atoms. Note that, typically, there exists an enhancement factor of several orders of magnitude between $E_0$ and $E_0^{i}$, i.e.~$E_0=\alpha E_0^{i}$. In our case, and considering an incoming laser intensity of about 10$^{11}$ W/cm$^{2}$, $\alpha\approx100$. 
%Note that this value corresponds to the plasmonic-enhanced one, not to the incoming intensity of the driving IR field, that could be several orders of magnitude less (see e.g.~\cite{kim2008high}).
The parameters $\beta_{x}$ and $\beta_{y}$ 
characterises the strengths of the plasmonic field along the $x$- and $y$-directions, respectively. 
The spatial inhomogeneity $h$ is typically approximated to linear order,
i.e., $h(x)=x$ and $h(y) = y$~\cite{husakou2011theory, ciappina2012high, yavuz2012generation, ansari2018simultaneous},
 which is the chosen form in our model too, and gives the dimension of inverse length to $\beta_{x/y}$. As it is discussed in Ref.~\cite{ciappina2012high}, the inverse of $\beta$ defines the spatial inhomogeneity region, i.e.~the spatial region where the field present noticeable spatial variations. For instance, a $\beta=0.01$ a.u corresponds to an inhomogeneity region of 100 a.u. (5.3 nm) (see next Section). This number appears to be compatible with, for instance, a distance between the bow-ties element apexes of around 10 nm.
The values of $\beta_{x}$ and $\beta_{y}$ can be tuned independently, for instance, changing the distance between the bowties apexes in either the $x$ or $y$ direction. 
Typically, if one chose the plasmonic nanostructure to be resonant with the incoming driving frequency, the plasmonic-enhanced field is an 'amplified' copy of the incoming one, with a maximum enhancement factor. As the plasmonic field development is not instantaneous, some phase difference between the incoming and outgoing fields can be observed~\cite{kling2027ieee}. For long pulses, as the ones used in this work, this phase difference is irrelevant. For the case of bichromatic driving fields, where two frequencies come into play, it would be more difficult to find the resonance condition, engineering the plasmonic nanostructures. As we are not interested in maximizing the amplification factor, this is not an issue. In principle, different time delays will appear for each of the two colors, but nowadays there exists enough time resolution as to control the incoming field in a way to compensate any phase difference inherited in the plasmonic-enhanced field.

From an experimental viewpoint, the seminal experiment of Kim {\it et al.}~\cite{kim2008high} was put under controversy several years after~\cite{sivis2012Nature}. In short, one of the problematic points was indeed the conversion efficiency of the nanostructure-enhanced HHG. Meanwhile Sivis et al~\cite{sivis2012Nature}, argued that the ratio between nanostructure-enhanced and conventional HHG is $\lesssim10^{-8}$, Kim {\it et al.}~\cite{kim2012Nature}, concluded that the ratio reached $10^{-6}$. The latter is obtained using a total number of atoms equal to $\approx 8\times10^{4}$ atoms, in a total interaction volume of 60 nm x 50 nm x 50 nm x 150 bow-ties at 115 torr pressure (for more details see Ref.~\cite{kim2012Nature}). Another point to take into account is that the signal could contain atomic lines emission, that could prevent to observe a `clean' HHG signal~\cite{sivis2013extreme}. We are aware that the extraction of a measurable and `bright' HHG signal from the original Kim's setup could be demanding, but, as it is discussed in~\cite{sivis2012Nature,kim2012Nature} and presented in~\cite{park2011plasmonic}, there exist another setups, where it would be possible to extract a more realistic number of HHG photons. From a theoretical viewpoint, however, we only would need to deal with a spatial inhomogeneous field, generated with whatever nanostructure setup, and using any target as a driven media. For instance, it is nowadays well established that the interband HHG in solid materials resembles to a great extent the atomic/molecular HHG. Therefore, if we consider a setup as the one presented in Ref.~\cite{han2016high}, it would be possible to apply our proposal to a piece of solid, embedded in a plasmonic nanostructure. Here, several problems are solved, namely (i) there is no risk to observe atomic lines emission and (ii) even when the interaction volume is comparable to the previous (atomic) case, the density of the solid would make the HHG signal larger.
 
 A final remark about our single atom model. It is true that any displacement of the atom in the interaction region will change the modulations and the structure of the HHG emission. There is, however, one point to consider. A measurable HHG signal will be only possible if the plasmonic-enhanced field exceed the value needed for the HHG process to happen. In this way, not all the atoms will `feel' an electric field able to generate HHG. This last aspect could enter in contradiction with some of the previous points, i.e.~we are reducing the interaction volume (number of atoms) even more, what would make the measurement of an HHG signal even more arduous. On the other hand, it is not the intention of the present work to cover both the microscopic and macroscopic/collective aspects (coherence of the emitted radiation, phase matching, etc.) of the process, but to present a `proof of principle', in the sense to show and study, at a single atom level, how a plasmonic-enhanced 2D field is able to control the polarization of the generated attosecond pulses.
 
\begin{figure*}
\includegraphics[width=\textwidth]{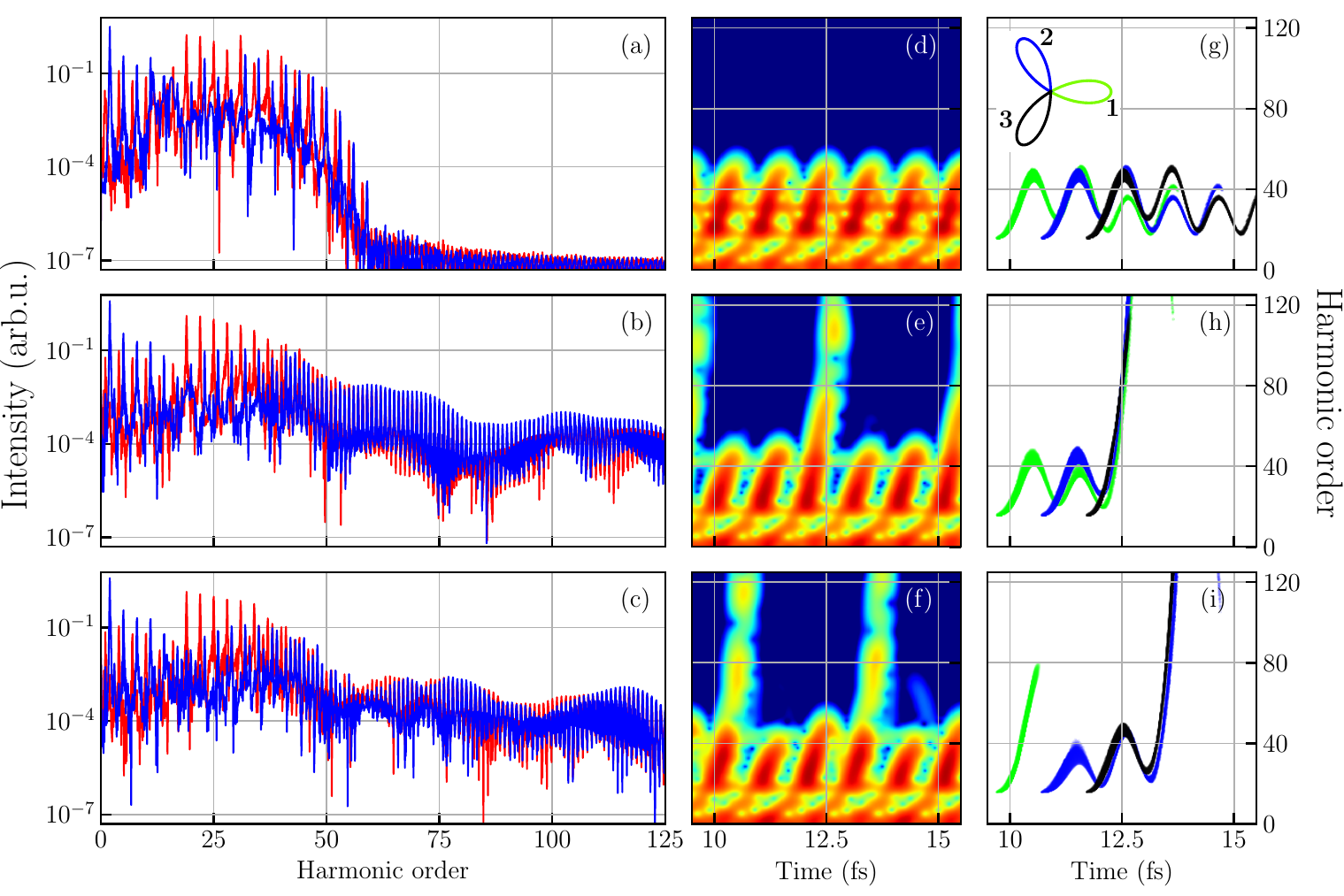}
\caption{High-order harmonic spectra (a-c) and time-frequency maps (d-f) calculated by solving 2D-TDSE. 
(g-i) represent the return-energy versus -time of the ionized electrons calculated from classical trajectory Monte-Carlo simulations. The harmonics plotted in red (or blue) colour in Fig. (a-c) are right- (or left-) circularly polarized. 
The Lissajous figure of the total bichromatic counter-rotating circularly polarised  driving field is trefoil and shown in the inset. The field lobes are marked numerically as they are traversed in time, i.e., in time, the field follows the order 1-2-3 of the lobes. Also, the field lobes are colour-coded with respect to their ionization and recombination times. 
In other words, the return-energy and -time of the electron which was ionized during, say, lobe 1 (green) is represented by green color in Figs.~(g-i). Moreover, the first row [Figs.~(a), (d), and (g)] is obtained for the spatially homogeneous field, i.e., $\beta_{x} = \beta_{y} = 0$. The second [Figs.~(b), (e), and (h)] and third [Figs.~(c), (f), and (i)] rows correspond to the cases when the plasmonic field is applied  along x-axis ($\beta_{x} = 0.01$ a.u.; $\beta_{y} = 0$) and along y-axis ($\beta_{x}=0;$ $\beta_{y} = 0.01$ a.u.), respectively (see the text for more details).} \label{fig2}
\end{figure*}

\section{Results}
 
Figure~\ref{fig2}(a) shows the   
reference harmonic spectrum obtained for the spatial homogeneous case with $\mathbf{E}^{\mathrm{pl}} = 0$.
To obtain the resultant harmonic spectrum, the electron is assumed to be emitted from the two degenerate 
atomic $p$-states, i.e., $p+$ ($l=1, m=+1$) and $p-$ ($l=1, m=-1$) and then, both the contributions are added coherently. 
Also, the time-dependent dipole was converted to a right/left circularly polarised basis by using 
$d_{\pm}=[d_{x}(t) \pm i d_{y}(t)]/ \sqrt{2}$, where $d_{x/y}(t)$ are the net time-dependent dipoles 
along $x/y-$coordinates. 
The obtained reference spectrum consists of  
$3n+1$ and $3n+2$ harmonics exhibiting right (red) and left-handed (blue)
circular polarisations, respectively, and the $3n$ harmonics are absent. 
This was studied in detail in previous works \cite{medivsauskas2015generating, jimenez2017time, milovsevic2015high, neufeld2019floquet, misha2018}.
The reference spectrum also shows
a clear energy cutoff at around $\approx50^{\textrm{th}}$ harmonic order,
which is consistent with the cutoff law for \BiCRCP fields of equal intensity~\cite{milovsevic2000generation}.  
Elliptically polarised attosecond pulses are expected since 
the spectral intensities of harmonics with opposite helicity
are different~\cite{medivsauskas2015generating}.

Figures~\ref{fig2}(b) and (c) present plasmonic field assisted harmonic spectra with  
$\beta_x = 0.01$ a.u. ($\beta_y= 0$) and $\beta_y = 0.01$ a.u. ($\beta_x = 0$), respectively. 
The spectra are drastically different in comparison to the reference case of Fig.~\ref{fig2}(a). 
First, there is a significant increase in the energy cutoff, which is a general feature of plasmonic-enhanced HHG \cite{ciappina2012high, ciappina2012enhancement, 
shaaran2012quantum,    yavuz2012generation, yavuz2013gas, shaaran2013quantum,  
shaaran2013high, perez2013beyond,  
fetic2013high, he2013wavelength, 
ebadi2014interferences, ciappina2014high,    luo2014efficient, chacon2015numerical,  feng2015attosecond,  ansari2018simultaneous}. 
Second, harmonics of order $3n$ appear in the spectra for higher orders, which 
is also expected, since the spatially-dependent plasmonic field $\mathbf{E}^{\mathrm{pl}}$ 
breaks the trefoil symmetry of the incoming field $\mathbf{E}^{\mathrm{total}}$. 
Furthermore, above $~35^{\textrm{th}}$ order 
the right- and left-circularly polarized harmonics overlap with approximately equal intensities
and have no apparent selection rules.
This is a strong indication that the harmonics above the $35^{\textrm{th}}$ order
originate from linearly or low-elliptically polarised attosecond pulses
generated once per laser cycle.
Note, that the differences in the spectra 
are sensitive to the direction of the plasmonic field since the 
spectra in Figs.~\ref{fig2}(b) and (c) are different.  

To connect the changes of the harmonic spectra with the underlying sub-cycle electron dynamics, 
we obtain time-frequency maps by performing a Gabor transform of the time-dependent total dipole 
\begin{equation}\label{gabor}
\tilde{d}_{x,y}(\Omega, t_{0}) = \frac{1}{2\pi}\int \mathrm{d}t\ d_{x,y}(t)~e^{-i\Omega t}e^{-\frac{(t-t_{0})^{2}}{(2\sigma^{2})}}
\end{equation}
with $\sigma=1/(3\omega)$. 
In the time-frequency map of the reference spectrum, 
the pair of short and long trajectories can be seen, which resemble the rising and falling edges of a gaussian curve, 
respectively. It is evident that the short trajectories are dominant. 
The reason can be found by analysing the harmonic generation process via semiclassical three-step model. 
The model gives a complex travel time of the electron in continuum owing to its generation by  tunnelling. 
The imaginary part of the travel time gives the probability of the process whereas the real part decides the energy of the harmonic. Milo{\v{s}}evi{\'c} and co-workers~\cite{milovsevic2000generation} have showed that 
the imaginary part peaks at approximately $T/3$ and drops substantially after half-a-cycle 
of the fundamental field  and thus, explained the major contribution of the harmonics produced by the short trajectory electrons. Also, it can be 
seen that three bursts of radiation of equal strengths (originated from short trajectories) are produced during each cycle $T \approx 2.6$ fs of the fundamental field (see Fig.~\ref{fig2}(d)). 
It can be understood by the Lissajous figure of the total field, 
which shows that the total electric field reaches its maximum three times during a single cycle of the fundamental field (trefoil symmetry; see inset in Fig.~\ref{fig2}(g)).

To further interpret the continuum electron dynamics and
underlying HHG process, we use 
classical trajectory Monte Carlo (CTMC) simulations.
Details of the CTMC calculation can be found in  \cite{Hofmann2018} and references therein. In addition to an ADK initial transverse momentum distribution \cite{ADK1986, Delone1991}, a Gaussian distribution of initial longitudinal momentum was allowed \cite{Hofmann2013, Hofmann2014}. 

The return-times and -energies of the ionized electrons in the reference driving field
obtained from the CTMC calculations are shown in Fig.~\ref{fig2}(g). 
The return condition for each trajectory was a shorter distance to the ion than the individual exit radius, 
which is known to reproduce TDSE and experimental results~\cite{Xiong2015, Hofmann2018}.
The CTMC trajectories are grouped according to the ionisation events occurring during one of the three lobes of the driving field,
indicated by different colors in Fig.~\ref{fig2}(g). Comparing Figs.~\ref{fig2}(d) and (g) we see that the CTMC calculations reproduce the Gabor analysis well. Furthermore, it is made clear that the main features in the reference spectra are produced by the short electron trajectories, i.e., the electrons that return during the next lobe of the field after their ionization.
This is in good agreement with trajectory-based studies of HHG in bi-chromatic counter-rotating field \cite{zhang2016helicity}.

%
% START DISCUSSING THE PLASMONIC FIELD EFFECTS HERE
%

As we have seen in Figs.~\ref{fig2}(a-b), the plasmonic field tends to break the symmetry of the harmonic spectra. In contrast, the time-frequency maps in Figs.~\ref{fig2}(d-e) reveal a very regular sub-cycle dynamics. The effect of the (symmetry-breaking) plasmonic field 
is to enhance (or suppress) the energy-cutoff of attosecond bursts of radiation produced during each lobe of the driving laser field. 
When the plasmonic field is along the $x$-direction (parallel to one of the field maxima directions) in Figs.~\ref{fig2}(b),(e) and (h),
those short trajectories which are ionised during a lobe immediately before the laser field aligns with the plasmonic field have enhanced kinetic energy (say, the trajectory shown in black color in Fig.~\ref{fig2}(h)). 
 Although the long trajectories does show an enhancement in harmonic energy, they have minimal probability leading to their negligible contribution in harmonic spectra.
The picture changes completely when the plasmonic field is oriented along $y$-direction (30$^{\circ}$ from the nearest field maximum) in Figs.~\ref{fig2}(c),(f) and (i). Here, the short trajectories ionised during the second lobe of the field return with reduced energy, while short trajectories ionised during the first lobe and long trajectories from both others contribute to the enhanced harmonics.

\begin{figure}[h]
	\includegraphics[width= 8.5 cm]{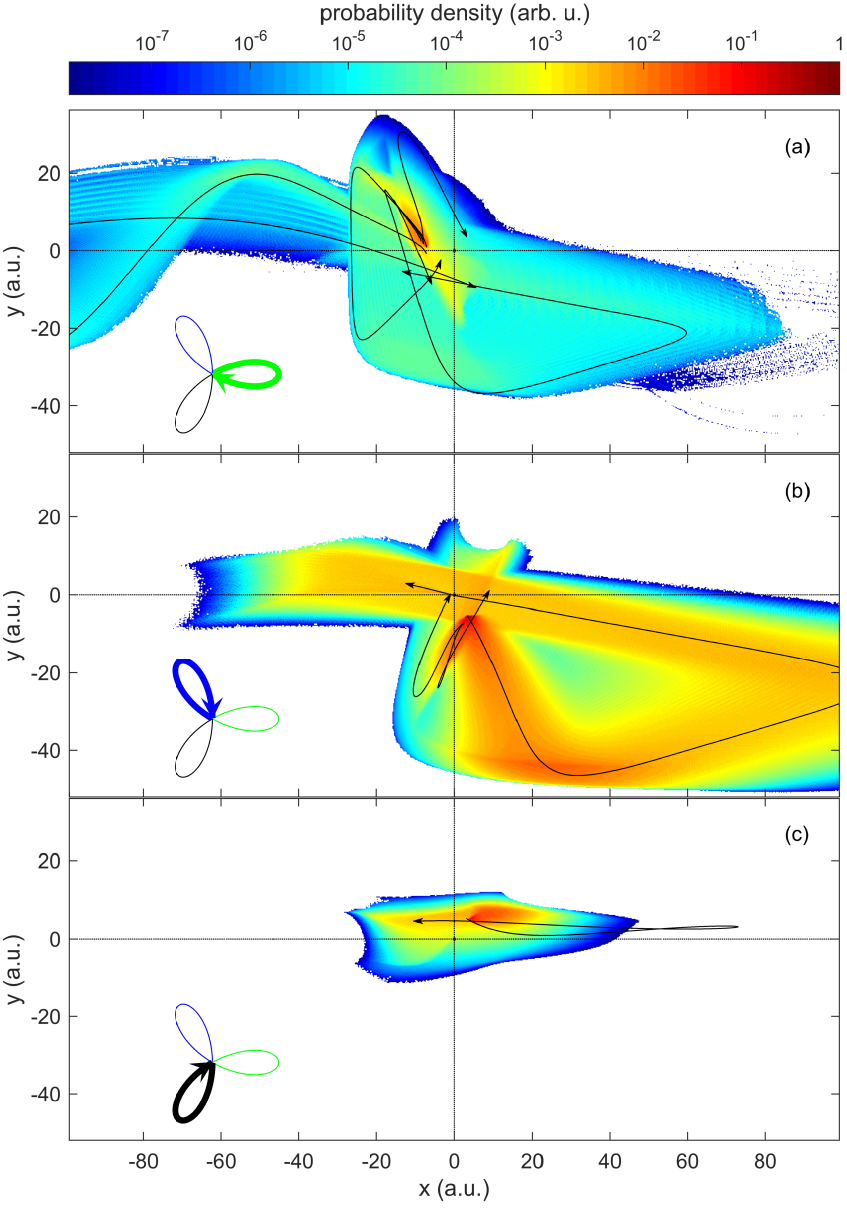}
	\caption{Classical electron trajectories, which are returning back to the parent ion, ionised during three different lobes of the field. The plasmonic field is employed along $x$-direction with $\beta_{x}$ = 0.01 a.u. Some exemplary trajectories are highlighted by black arrows. The insets indicate the lobe during which the ionization takes place and the rotation direction of the laser electric field vector.
	} 
	\label{fig3}
\end{figure}

The enhancement of attosecond radiation bursts becomes clear, when looking at the
probability density maps of returning trajectories in Fig.~\ref{fig3}. 
They are obtained by CTMC simulation, for ionisation during each lobe of the total field
and for the plasmonic field applied along $x$-direction ($\beta_{x}$ = 0.01 a.u.). 
If the electron is ionized during the lobes 1 or 2, 
the excursion of its returning trajectory can be much further 
and these electrons can recombine during the next several lobes after ionization 
[see Figs.~\ref{fig3}(a) and (b)].  
Contrarily, only short trajectories return  
when an electron is ionised during lobe 3 and recombines 
during lobe 1 of the next cycle [see Fig.~\ref{fig3}(c)].

Trajectories returning towards the ion from the positive $x$-axis are those which yield harmonics with enhanced cut-off energy [compare Fig.~\ref{fig3} and Fig.~\ref{fig2}(h)]. 
In this case, both the direction of the applied plasmonic field and the direction of the driving field coincides with the direction of returning photoelectron such that the electron feels the maximum acceleration upon return and accumulates more kinetic energy. 
Similar reasoning can be used if we apply the plasmonic field along $y$-direction 
($\beta_{y}$ = 0.01 a.u.). 
In this case, higher energy photons are produced when photoelectrons
recombine during lobe 2 of the field, see again Fig.~\ref{fig2}(i).
If we reverse the helicity of the laser field, i.e., choosing negative helicity for the $\omega-$field and positive helicity for the $2\omega-$field, the order of the lobes followed by the net bicircular field will change. It means that instead of lobes 1-2-3, the order becomes 1-3-2 in time. By considering the reverse helicity field, while keeping $\beta_y = 0.01$~a.u., the short trajectories returning at approximately $12$ fs will be enhanced because they are ionised when the field points mostly along negative $y$-direction and return from positive $y$-direction during the next lobe (not shown here). This is different than the case shown in Figs.~\ref{fig2}(b), (e) and (h) when $\beta_{x} = 0.01$ lead to the enhancement of trajectory returning at $12$ fs.

%
% START DISCUSSING STRONGER PLASMONIC FIELDS
%

\vspace{0.2in}

\begin{figure}[t]
\includegraphics[width = 12 cm]{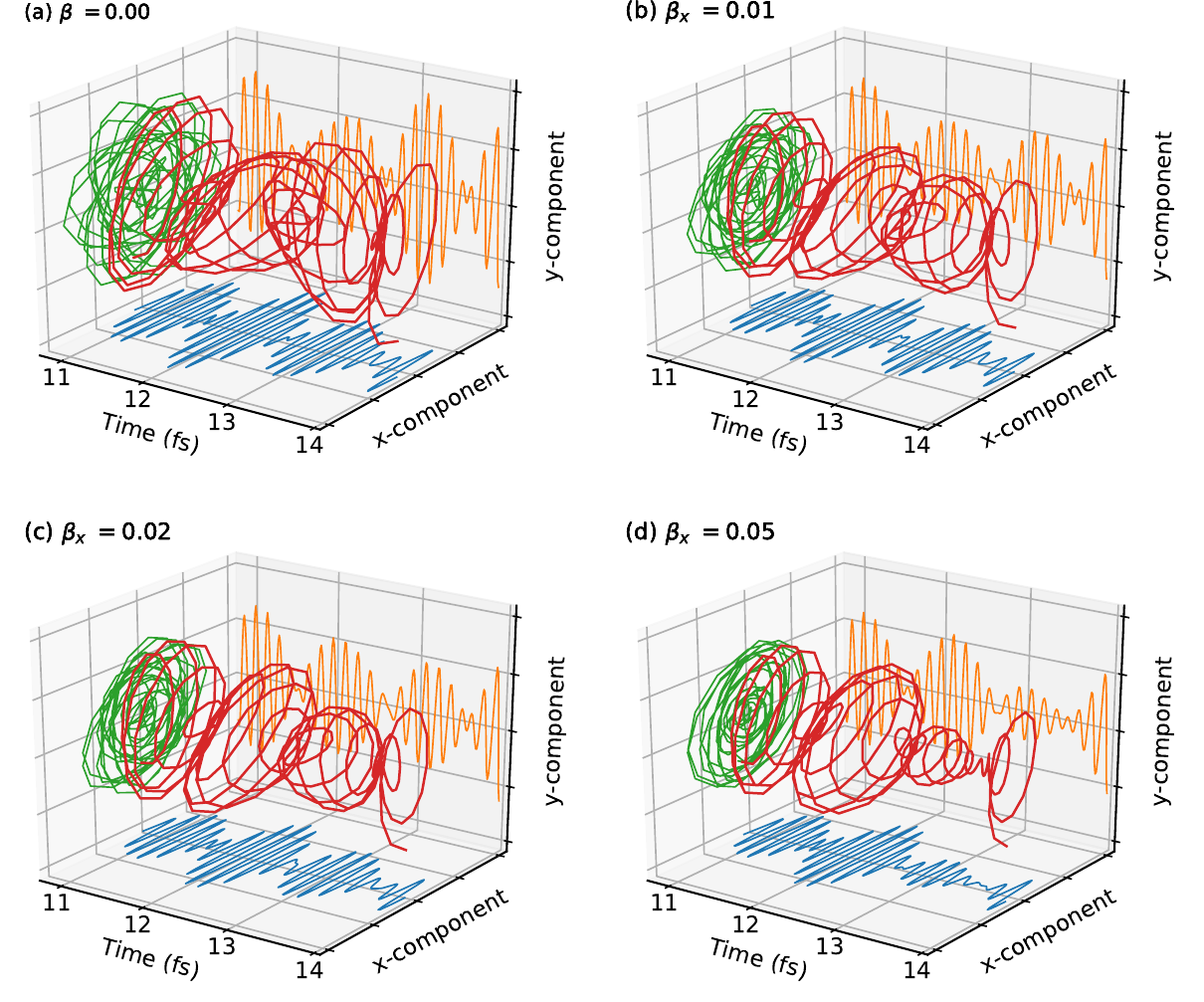}
\caption{The total time-dependent electric field (in red) corresponding to an attosecond pulse train for different strength of plasmonic field along $x$-direction.  
The x-component (in blue),  y-component  (in orange) and Lissajous figure (in green) of the total field are also shown.
$18-24^{\textrm{th}}$ harmonics window is used to synthesis an attosecond pulse train. } 
\label{fig4}
\end{figure} 

So far we have discussed only the results for a plasmonic field with strength $\beta_{x,y} = 0.01$ a.u.,
for which the polarisation and position of the harmonics below the cut-off energy remain largely unaffected. 
As the strength of the plasmonic field is increased to $\beta_{x,y}$ = 0.02 and 0.05 a.u., 
the below-cutoff harmonic spectrum undergoes a substantial change with the appearance of 
$3n$ harmonics, and right- and left-handed harmonics overlapping with equal strength.
This modification is driven by the change of polarisation of the underlying attosecond pulses.
We demonstrate this change by plotting, in Fig.~\ref{fig4}, the total electric field and its $x$- and $y$-components
for a window between the $18^\textrm{th}-24^{\textrm{th}}$ harmonics
and for different plasmonic field strengths.

For the homogeneous case in Fig.~\ref{fig4}(a), the attosecond pulses have elliptical polarization,
where the major polarization axis is rotated by 120$^{\circ}$ for each pulse. This is a direct consequence of unequal intensities of the $3n+1$ and $3n+2$ harmonics and the trefoil symmetry of the BiCRCP field
\cite{medivsauskas2015generating, jimenez2017time, dorney2017helicity, neufeld2018optical, milovsevic2015high}. 
Indeed, if we look at the electric field generated by the harmonics around the harmonic cut-off energy, where both $3n+1$ and $3n+2$ harmonic intensities are similar [see Fig.~\ref{fig2}(a)], the attosecond pulses would show a near-linear polarization with the polarization direction rotated by 120$^{\circ}$ for each consecutive pulse.

As the strength of the plasmonic field increases, the polarization of one radiation burst (emitted at around 13~fs in Fig.~\ref{fig4}), changes to circular and its amplitude is suppressed.
Meanwhile, the major polarization axis of the other two radiation bursts aligns to a common direction.
When the plasmonic field is applied along the $y$-direction (not shown here), analogous findings can be applied, 
with the difference that: (i) a  weaker plasmonic field ($\beta_{y}=0.01$ a.u.) is already sufficient to align the polarization ellipses of the generated attosecond pulses and (ii) suppression of one of the radiation bursts occurs only for stronger field strengths.

Beyond the harmonic cut-off energy, 
the emitted radiation has linear polarization 
with a direction that coincides with that of the applied plasmonic field.
Consequently, all the harmonics including $3n$-harmonics have comparable intensities.  
For example, the electric field obtained by filtering out $60^{\textrm{th}}-66^{\textrm{th}}$ harmonics
has a linear polarization along $x$- ($y$-) direction  when $\beta_{x} \neq 0.00$ 
($\beta_{y} \neq 0.00$).  
The same is true for the electric field generated by $78^{\textrm{th}}-84^{\textrm{th}}$ harmonics. 

This observation corresponds to what we have already seen from the CTMC analysis. 
The HHG spectrum above the reference cut-off energy 
stems from the photoelectrons that return to the parent ion from the direction of the plasmonic field
and therefore, are additionally accelerated to higher kinetic energy.
However, since the additional acceleration occurs only along the plasmonic field direction,
the high-energy radiation is linearly polarized and is emitted only once per laser cycle.
Any photoelectron trajectories returning and recombining from a different angle will only contribute to the lower energy part of the HHG spectrum.

\section{Conclusions}
In summary, while the HHG process driven by \BiCRCP laser field follows the three-fold symmetry pattern, adding a local plasmonic field along a particular spatial dimension modifies the sub-cycle electron dynamics in non-trivial ways: enhancing some long or short photoelectron trajectories while suppressing some others. Simply by changing the direction of plasmonic field enhancement, rich yet predictable control of electron trajectories is possible, providing as yet unexplored possibilities for sub-cycle control of the HHG process. Furthermore, the modification of the sub-cycle electron dynamics also leads to subtle changes in the polarization of the emitted high-harmonic radiation. For instance, by filtering appropriate harmonic orders, an attosecond pulse train consisting of nearly-circularly, elliptically or linearly polarised pulses is achieved. Alternatively, by changing the plasmonic field intensity, a predefined major axis direction is enforced for all elliptically polarized pulses in an attosecond pulse train. The plasmonic field intensity can be tuned by geometrically engineering plasmonic nanostructures with different sizes and shapes.

\acknowledgments

G. D. acknowledges support from Science and Engineering Research Board (SERB) India 
(Project No. ECR/2017/001460) and 
Max-Planck India visiting fellowship.  We acknowledge support from ERC AdG NOQIA, Spanish Ministry of Economy and Competitiveness (``Severo Ochoa" program for Centres of Excellence in R\&D (CEX2019-000910-S), Plan National FISICATEAMO and FIDEUA PID2019-106901GB-I00/10.13039 / 501100011033, FPI), Fundaci\'o Privada Cellex, Fundaci\'o Mir-Puig, and from Generalitat de Catalunya (AGAUR Grant No. 2017 SGR 1341, CERCA program, QuantumCAT\_U16-011424, co-funded by ERDF Operational Program of Catalonia 2014-2020), MINECO-EU QUANTERA MAQS (funded by State Research Agency (AEI) PCI2019-111828-2/10.13039/501100011033), EU Horizon 2020 FET-OPEN OPTOLogic (Grant No 899794), and the National Science Centre, Poland-Symfonia Grant No. 2016/20/W/ST4/00314. MFC receives support from the Grantov\'a agentura \u{C}esk\'e Republiky (GA\u{C}R Grant 20-24805J).

%\bibliography{references}

%merlin.mbs apsrev4-1.bst 2010-07-25 4.21a (PWD, AO, DPC) hacked
%Control: key (0)
%Control: author (8) initials jnrlst
%Control: editor formatted (1) identically to author
%Control: production of article title (-1) disabled
%Control: page (0) single
%Control: year (1) truncated
%Control: production of eprint (0) enabled
%

\end{document}